\documentclass[a4paper, amsfonts, amssymb, amsmath, reprint, superscriptaddress, prb, aps]{revtex4-2}
\usepackage[english]{babel}
\usepackage[utf8]{inputenc}
\usepackage[colorinlistoftodos, color=green!40, prependcaption]{todonotes}

\usepackage{amsthm}
\usepackage{mathtools}
\usepackage{physics}
\usepackage{xcolor}
\usepackage{graphicx}
\usepackage[left=23mm,right=13mm,top=35mm,columnsep=15pt]{geometry} 
\usepackage{adjustbox}
\usepackage{placeins}
\usepackage[T1]{fontenc}
\usepackage{lipsum}
\usepackage{csquotes}
\usepackage{soul}
\usepackage{float}
\usepackage{hyperref} % For hyperlinks in the PDF
\begin{document}
\title{Nonvolatile Cryogenic Phase Slip Memory with Single-Shot Readout}

\author{Lukas Nulens}
    \email{lukas.nulens@kuleuven.be}
    \affiliation{Quantum Solid-State Physics, Department of Physics and Astronomy, KU Leuven, Celestijnenlaan 200D, Leuven, B-3001, Belgium}

\author{Davi A. D. Chaves}
    \affiliation{Quantum Solid-State Physics, Department of Physics and Astronomy, KU Leuven, Celestijnenlaan 200D, Leuven, B-3001, Belgium}

\author{Stijn Reniers}
    \affiliation{Quantum Solid-State Physics, Department of Physics and Astronomy, KU Leuven, Celestijnenlaan 200D, Leuven, B-3001, Belgium}

\author{Ruben Dillemans}
    \affiliation{Quantum Solid-State Physics, Department of Physics and Astronomy, KU Leuven, Celestijnenlaan 200D, Leuven, B-3001, Belgium}

\author{Ivo P. C. Cools}
    \affiliation{Quantum Device Physics Laboratory,
Department of Microtechnology and Nanoscience,
Chalmers University of Technology, Kemivägen 9, Goteborg, SE-412 58, Sweden}

\author{Kristiaan Temst}
    \affiliation{Quantum Solid-State Physics, Department of Physics and Astronomy, KU Leuven, Celestijnenlaan 200D, Leuven, B-3001, Belgium}
    \affiliation{Imec, Kapeldreef 75, 3001 Leuven, Belgium}

\author{Bart Raes}
    \affiliation{Imec, Kapeldreef 75, 3001 Leuven, Belgium}

\author{Margriet J. Van Bael}
    \affiliation{Quantum Solid-State Physics, Department of Physics and Astronomy, KU Leuven, Celestijnenlaan 200D, Leuven, B-3001, Belgium}

\author{Joris Van de Vondel}
    \email{joris.vandevondel@kuleuven.be}
    \affiliation{Quantum Solid-State Physics, Department of Physics and Astronomy, KU Leuven, Celestijnenlaan 200D, Leuven, B-3001, Belgium}

\date{\today} 

\begin{abstract}
The demand for cryogenic memory components is driven by the need for ultra-fast, low-power, and highly reliable computing systems. Phase slip-based devices promise to fulfill all these requirements, with potential applications in both classical and quantum information processing. However, previous implementations have faced challenges due to inefficient writing and readout schemes. In this work, we address these limitations with a simple device design and measurement techniques inspired by circuit quantum electrodynamics. We present a memory element that stores information in the winding of a high-kinetic inductance superconducting loop, inductively coupled to a coplanar waveguide resonator. Using single-shot measurements, we achieve a readout fidelity of 99.698\% with an active measurement time of just 25 ns.

\end{abstract}
%\keywords{first keyword, second keyword, third keyword}

\maketitle

\section*{Introduction}

Quantum technologies are moving from academic research to applications in various areas of life \cite{kjaergaard2020superconducting,bayerstadler2021industry,krelina2021quantum,bravyi2022future,luo2023recent,flother2023how}. This progress has been driven both by a better understanding of quantum phenomena and by innovations in nanofabrication, microwave circuitry, and cryogenic technologies \cite{biswas2012advances,Blais2021circuit,zu2022development,zwerver2022qubits,Burkard2023semiconductor,van2024advanced,borsoi2024shared}. As technology advances, developing new circuit elements compatible with the demands of modern computing systems has become crucial. These systems typically operate in sub-Kelvin environments, utilizing radiofrequency manipulation and readout in the 4 to 8 GHz range for superconducting qubits. Superconducting elements naturally emerge as strong candidates for integration, thanks to their low-dissipation even for high frequency signals. Previous efforts have led to the development of various new superconducting circuit elements, including diodes \cite{vandevondel2005vortex,de2006controlled,ando2020observation,lin2022zero,nadeem2023superconducting}, parametric amplifiers \cite{castellanos2007widely,ho2012wideband}, and single photon detectors \cite{natarajan2012superconducting,esmaeil2021superconducting}.

%Particularly, developing reliable and efficient memory elements to enable classical bit operations in cryogenic environments is an important task.
The potential of superconductors for fast and energy efficient classical computing tasks is particularly demonstrated by the development of single-flux-quantum (SFQ) logic since the 1970s, using Josephson junctions as fundamental circuit elements \cite{likharev1991rsfq,mukhanov2011energy,holmes2013energy}.
%During the 1980s, rapid single-flux-quantum (SFQ) logic was developed and has since been optimized, enabling fast classical information processing using Josephson junctions as the fundamental circuit elements \cite{likharev1991rsfq,mukhanov2011energy,holmes2013energy}. 
Recently, SFQ logic has been used to control quantum systems \cite{McDermott2018quantum,Liu2013single}. 
%However, although fully-superconducting processors exist, they have yet to replace their semiconducting counterparts. 
One significant challenge both to SFQ electronics and quantum computing is to design suitable memory elements to work in cryogenic environments, with different approaches still being explored \cite{alam2023cryogenic,Aziz2025}. Superconducting memory elements are attractive for their potential to operate at high speeds with minimal energy dissipation.
These include solutions based on vortex traps \cite{golod2023word,hovhannisyan2024controllable,foltyn2024quantum}, Josephson junctions \cite{chen2020miniaturization}, hybrid superconductors \cite{held2006superconducting,baek2014hybrid,zhao2018compact}, kinetic inductance \cite{chen1992kinetic,mccaughan2018kinetic}, and geometrical constrictions of superconducting materials \cite{kalashnikov2024demonstration,murphy2017nanoscale,ligato2021preliminary,ilin2021supercurrent,chaves2023nanobridge}.

The two latter approaches rely on fluxoid quantization in superconducting loops \cite{tinkham2004introduction}. This gives rise to a discrete set of solutions or energy states at each specific applied magnetic field. Each energy state corresponds to a winding number or vorticity, representing the number of enclosed flux quanta. A characteristic energy barrier avoids transitions between these states, assuring that both can survive under a particular magnetic field. This metastability is necessary for the phase-slip element 
%thus introducing the necessary metastability for the phase-slip element
to function as a nonvolatile classical memory. The energy barrier is sensitive to both applied currents and magnetic fields \cite{ligato2021preliminary,chaves2023nanobridge,lau2001quatum,nulens2022metastable}, which can trigger local fluctuations of the superconducting order parameter, known as phase slips, which change the superconducting phase in integer multiples of 2$\pi$\cite{lau2001quatum,mooij2006superconducting}. These transitions between different winding states \cite{Hazra2019nanobridge,nulens2022metastable,potter2023controllable,mooij2005phase} arise from a reduction of the energy barrier and a shift in the energy balance due to the external biases. Consequently, with precise control over the energy barrier, phase slips can be utilized to define a deterministic writing protocol. The barrier is also intrinsically linked to the kinetic inductance and geometry of the superconducting element. Specifically, it decreases as the loop’s cross-sectional area decreases \cite{mooij2005phase,dausy2021impact,peltonen2013coherent}. As such, a common strategy for controlling the barrier is to incorporate a weak link into the superconducting loop \cite{potter2023controllable,astafiev2012coherent,belkin2015formation}. This is important, because fabricating devices to work as classical memory elements requires a certain balance. 
The barrier must be designed such that it is large enough to protect the system from unwanted phase-slip events without requiring large excitations to switch between states. It should also not be too small, as metastability can be lost with the system always following the lowest energy state, potentially creating a phase-slip qubit \cite{mooij2005phase,astafiev2012coherent,purmessur2025operation}.

\begin{figure*}
    \centering
    \includegraphics[width=1\linewidth]{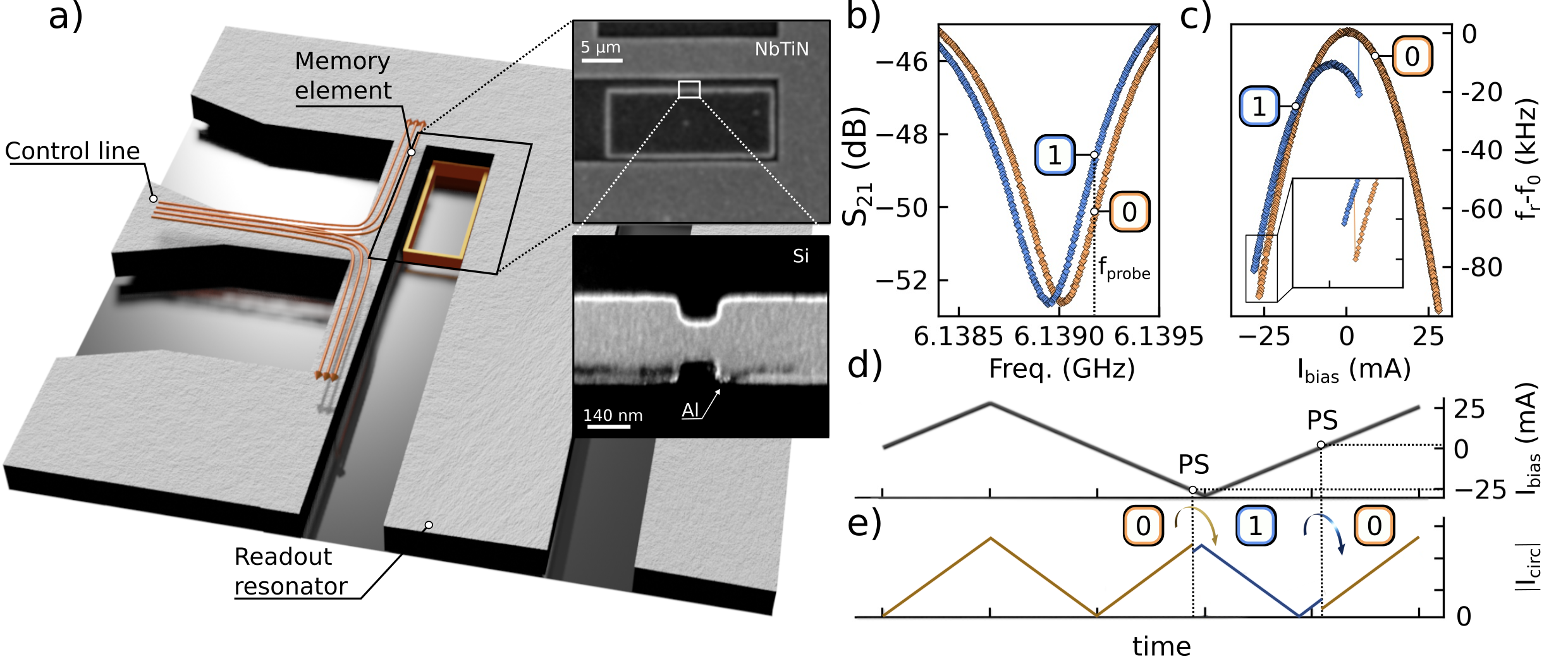}
    \caption{ a) A schematic representation of the memory device comprising an Al loop (indicated in orange) coupled to a NbTiN $\lambda$/4 resonator (light gray). A current bias line (control line) supplies a local magnetic field in the vicinity of the Al loop, which is positioned at the Si trench (dark gray). The orange arrows indicate the path of the bias current. A complete view of the device is found in the Supplementary Information. The insets show SEM images of the loop and constriction. b) Spectroscopy of the transmission through the feedline near the resonator's $f_r$ for two distinct energy states of the loop, labeled 0 and 1. c) The variation in resonance frequency following ZFC as $I_{bias}$ is altered following the values represented in panel d as a function of time. $f_0$ = 6.139074 GHz is the resonance frequency of state 0 at zero bias. Both phase slip events (from 0 to 1 and from 1 to 0) manifest as discontinuous jumps. The inset shows a magnified view of the 0 to 1 phase slip event. e) A schematic representation of the absolute value of the circulating current inside the Al loop during the $I_{bias}$ cycle. At the values indicated in panel d, a phase slip (PS) event occurs, altering the loop's energy state.}
    \label{fig:1}
\end{figure*}

Earlier phase-slip-based memory elements faced several challenges, with one of the main issues being their non-deterministic writing schemes that relied on the probability that the memory element would relax into an energy-favorable winding as the system transitioned from the normal to the superconducting state \cite{ilin2021supercurrent, chaves2023nanobridge}. Another challenge was the readout process, which involved either full critical current measurements or monitoring a non-zero DC voltage across the galvanically connected memory element \cite{ligato2021preliminary, ilin2021supercurrent, chaves2023nanobridge}. This makes both the read and write processes energy-inefficient and inherently slow, significantly limiting practical applications of these devices. In addition, some proposed devices also relied on external magnetic fields to bias the memory element and induce phase slips \cite{ligato2021preliminary, chaves2023nanobridge}, further complicating the integration of these memories into superconducting circuits due to the sensitivity of other on-chip components to magnetic fields. 

% \st{Combining the low-dissipation manipulation of classical phase-slip memories with well-established circuit quantum electrodynamics (cQED) measurement principles citeBlais2021circuit can dramatically increase the operation speed in these devices. Furthermore, the absence of external fields, along with a similar and straightforward design and fabrication process, as well as the use of GHz-range electronics, would make these memory elements compatible with current superconducting-based quantum technology.}

In this work, we demonstrate the operation of a phase-slip-based classical bit that addresses all of these concerns. Using a one-step fabrication approach, we create a superconducting loop containing a constriction-type junction, which tailors the device's characteristic energy barrier against phase slips. Building on recent advancements in on-chip inductive coupling of superconducting structures \cite{janvier2015coherent, nulens2024noninvasive, chiodi2011probing} and energy-efficient readout techniques \cite{gumucs2023calorimetry, kalashnikov2024demonstration}, including standard qubit single-shot readout \cite{kjaergaard2020superconducting, mallet2009single, Walter2017rapid}, we integrate the loop with a coplanar waveguide superconducting resonator. This configuration allows for reliable winding manipulation of nonvolatile states down to 13 mK, without the need for external magnetic fields.

\section*{Results}

The memory element in this work is a 15-nm-thick, 300-nm-wide high-kinetic-inductance Al loop, as shown by scanning electron microscopy (SEM) in the insets of Fig. \ref{fig:1}a\textemdash see the Supplementary Information for an overview of the design. The loop is fabricated using molecular beam epitaxy (MBE) without capping layer, allowing a native oxide layer of approximately 2 to 3 nm to protect it from further oxidation, as verified by XRR measurements of a simultaneously grown reference film. The MBE chamber maintained a base pressure better than 4$\times10^{-9}$ mbar, with a deposition rate of 1.3 \AA/s. The loop features a 143-nm-wide, 115-nm-long constriction, visible in the SEM image in the bottom inset of Fig. \ref{fig:1}a. 
%\st{This constriction allows control over the energy barrier protecting the loop against phase slips.}
This constriction defines the preferential location of an induced phase slip event, while its geometry can be used to control the height of the energy barrier. 
%\st{However, if the barrier is too high (i.e., the constriction has a large cross-sectional area), reliably switching the loop between energy states becomes difficult or requires large biases, which can compromise the readout circuitry. Conversely, if the barrier is too small, the system will always revert to the lowest energy state, becoming volatile and unsuitable as a classical memory element.} 
The constriction dimensions presented in this work are optimized to 
%\st{lie in an intermediary regime, in which phase slips are favored}
provide an efficient writing protocol while maintaining the nonvolatile behavior of the energy states.

The loop is coupled without any electrical contacts to a NbTiN $\lambda/4$ coplanar waveguide superconducting resonator. We model such a coupling by the mutual inductance of the loop-resonator system \cite{nulens2024noninvasive}. The resonator is capacitively coupled to a 50 $\Omega$ impedance-matched waveguide feedline and is grown on top of a high-resistance Si substrate. The coupling of the resonator and the loop causes the resonance frequency of the resonator ($f_r$) to depend on the superconducting properties of the Al loop, particularly the Cooper pair density ($n_s$) \cite{nulens2024noninvasive}. 
This behavior is confirmed by the temperature dependence of $f_r$, which reveals the transition of the loop to the superconducting state at 1.6 K (see Supplementary Information). The $f_r$ value is determined by monitoring the transmission ($S_{21}$) through the feedline and fitting the data using the method outlined in Ref. \cite{probst2015efficient}. Low-temperature radiofrequency characterization is performed using a Keysight P5003B Streamline Vector Network Analyzer (VNA) and a control system from Keysight comprised of an M5300A PXIe RF arbitrary waveform generator (AWG), a M5201A PXIe downconverter, and a M5200A PXIe digitizer.

As shown in Fig. \ref{fig:1}a, a current bias ($I_{bias}$) control line is positioned near the constriction, generating a local magnetic field that affects the loop's properties but not the readout resonator \cite{nulens2024noninvasive}. The magnetic field induces circulating shielding currents in the loop $I_{\text{circ}}$, which decrease both the energy barrier and $n_s$. As a result, a phase-slip event can occur, abruptly altering the winding, decreasing $I_{\text{circ}}$, and increasing $n_s$. At a fixed field value, different winding states correspond to different circulating currents and $n_s$ values, leading to distinct resonance frequencies. Thus the winding states can be probed by monitoring the resonator's resonance frequency. This is shown in Fig. \ref{fig:1}b, where the colored solid data points represent the transmission $S_{21}$ as a function of frequency near resonance for two distinct states, labeled 0 and 1. This readout mechanism is non-destructive, i.e., it does alter the winding state, and allows us to probe the memory state in an energy-efficient manner, 
%without disrupting \lukas{superconductivity}\st{its superconducting state} \lukas{not happy with that, give it a shot} or 
not relying on a dissipative voltage readout.

In Fig. \ref{fig:1}c, we track the change in $f_r$ relative to $f_0$ at 300 mK as $I_\text{bias}$ is swept from 0 to +28 mA and then back to $-28$ mA (orange data), with the $I_\text{bias}$ cycle schematically illustrated in Fig. \ref{fig:1}d. Here, $f_0 = 6.139074$ GHz represents the resonance frequency of the energetically preferred state with zero circulating current at zero magnetic field, which we define as energy state 0. The resulting $f_r$ parabola is centered at $I_\text{bias} = 0$, as expected after zero-field cooling (ZFC) \cite{nulens2022metastable}. At $I_\text{bias} = -26.45$ mA, a sharp discontinuity in $f_r$ is observed due to a change in $n_s$ caused by a phase slip, thus changing the winding of the loop. This state is characterized by a different $f_r(I_{bias})$ curve and will be referred to as state 1. Figure \ref{fig:1}e schematically demonstrates the evolution of the absolute value of $I_{circ}$ as $I_{bias}$ is altered, highlighting the change from state 0 to state 1 when a phase slip (PS) occurs. When $I_\text{bias}$ is subsequently increased to +28 mA (blue data), another phase slip occurs at +3.72 mA, returning the system to state 0.

Thus, when the loop is in state 0, applying an $I_\text{bias}$ pulse to $-28$ mA causes it to transition to state 1. In contrast, if the loop is in state 1, a pulse to +28 mA returns it to state 0. Despite the states' labels, the observed phase slips do not represent the change of a single winding in the loop. 
Once a phase slip occurs, a subtle balance between the amount of energy stored in the loop for a given winding, the energy barrier, and the phase dynamics enables phase changes of one or more integer multiples of 2$\pi$ \cite{nulens2022metastable}. As such, if a phase slip removes the loop from an energy state with a given winding $n$, the state with winding $n+1$ is not necessarily the lowest in energy at the biases where the switch occurs. For the device to function as a classical bit, nonetheless, what is crucial is that, after each write pulse, the loop deterministically switches between the same two states, which should be clearly distinguishable at zero bias current. In the following experiments, we test and confirm that the device presents these characteristics.

\begin{figure}
    \centering
    \includegraphics[width=1\linewidth]{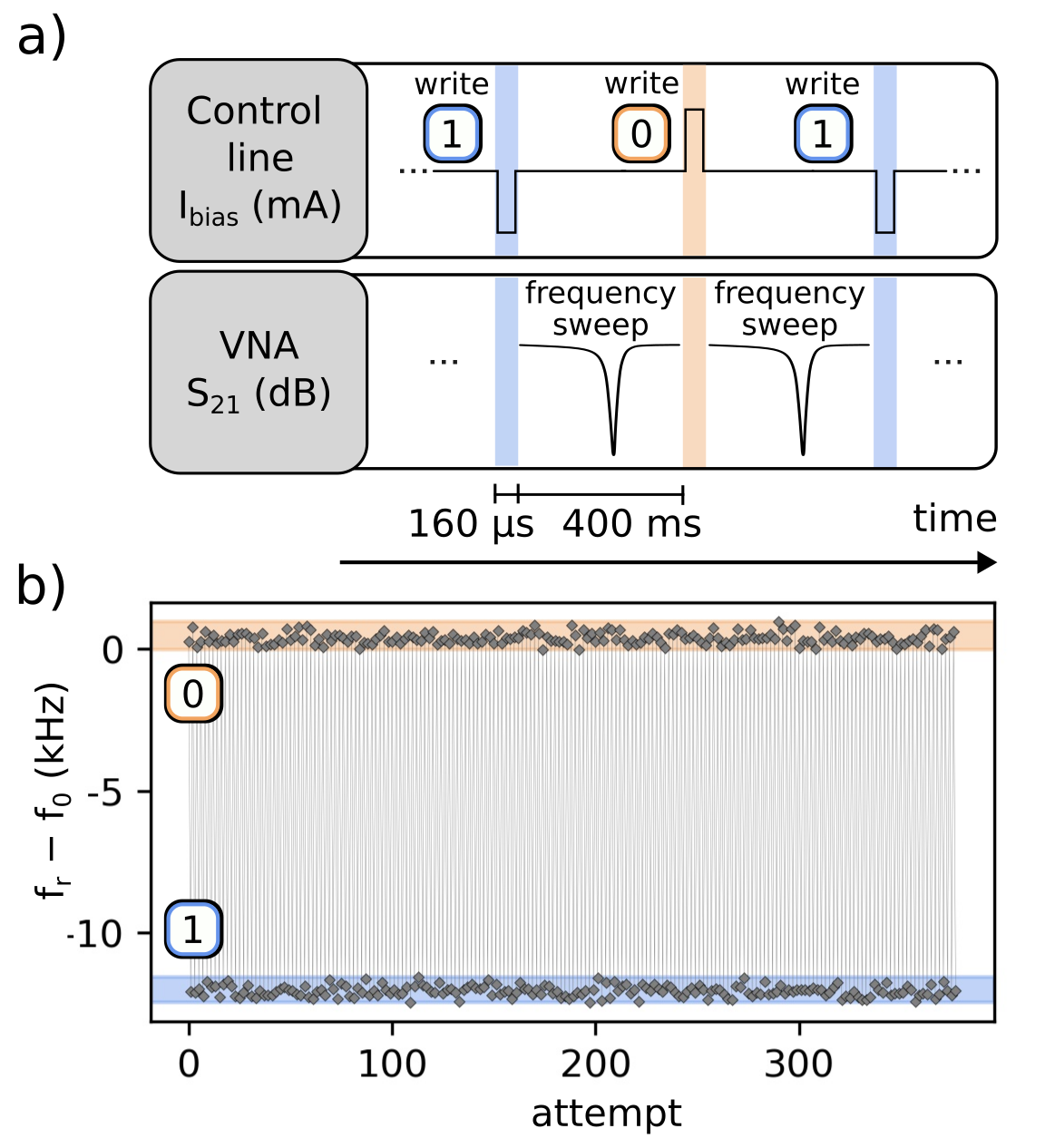}
    \caption{a) A schematic representation of the measurement protocol. After each write current pulse of $\pm 28$ mA, the VNA performs a frequency sweep. The resonance frequency is determined by fitting the obtained data, as described in the text.
    b)The resulting $f_r-f_0$ values at 300 mK, with $f_0$ = 6.139074 GHz, obtained after 375 alternating 160 $\mu$s write pulses. Orange and blue shaded regions correspond to states 0 and 1, respectively.}
    \label{fig:2}
\end{figure}

First, the stability of the memory states was confirmed by continuously monitoring $S_{21}$ for approximately five hours at zero bias (see Supplementary Information). Then, we validate the effectiveness of the writing principle for the memory element while performing frequency sweep readout measurements at 300 mK, as schematically represented in Fig. \ref{fig:2}a. Figure \ref{fig:2}b shows the results of a total of 375 $I_{\text{bias}}$ pulses, each lasting 160 $\mu$s and alternating between $\pm 28$ mA, applied to manipulate the state. Following each pulse, a complete frequency sweep is conducted at zero bias to determine $f_r$ using the VNA. The results confirm that the loop switches between the same two, clearly distinguishable states after the pulses, 
%\st{with no errors observed during these measurements.}
with no failed or incorrect transitions observed during the measurements. It is important to notice that the device is sensitive to fluctuations on the bias line, which can trigger unwanted transitions to different energy states. To mitigate this, we implemented low-pass $\pi$-filters with a 1 MHz cutoff frequency. By employing triangular-shaped pulses, which reduced current spikes associated to very small rise times in square-shapes pulses, we managed to optimize the writing speed to 20-$\mu$s. This speed was only limited by the employed electronics, wiring, and connectors, and does not represent the fastest achievable memory operation. Based on phase-slip dynamics, switching speeds on the order of picoseconds are possible \cite{likharev1991rsfq}, with the choice of materials impacting the dynamics characteristic times \cite{Vodolazov2007rearrangement,vandevondel2011vortex}. If such a fast switching is achieved, phase slip memories could be competitive with the fastest currently available cryogenic memory alternatives \cite{alam2023cryogenic,Aziz2025}. After optimization, the reliability of the write operation was further assessed by performing 6000 consecutive write pulses, observing no errors, as shown in the Supplementary Information. Due to the technical limitations in the writing speeds, we refrained from a complete analysis of the writing efficiency, and turned our attention to the readout of the memory states.

\begin{figure}
    \centering
    \includegraphics[width=1\linewidth]{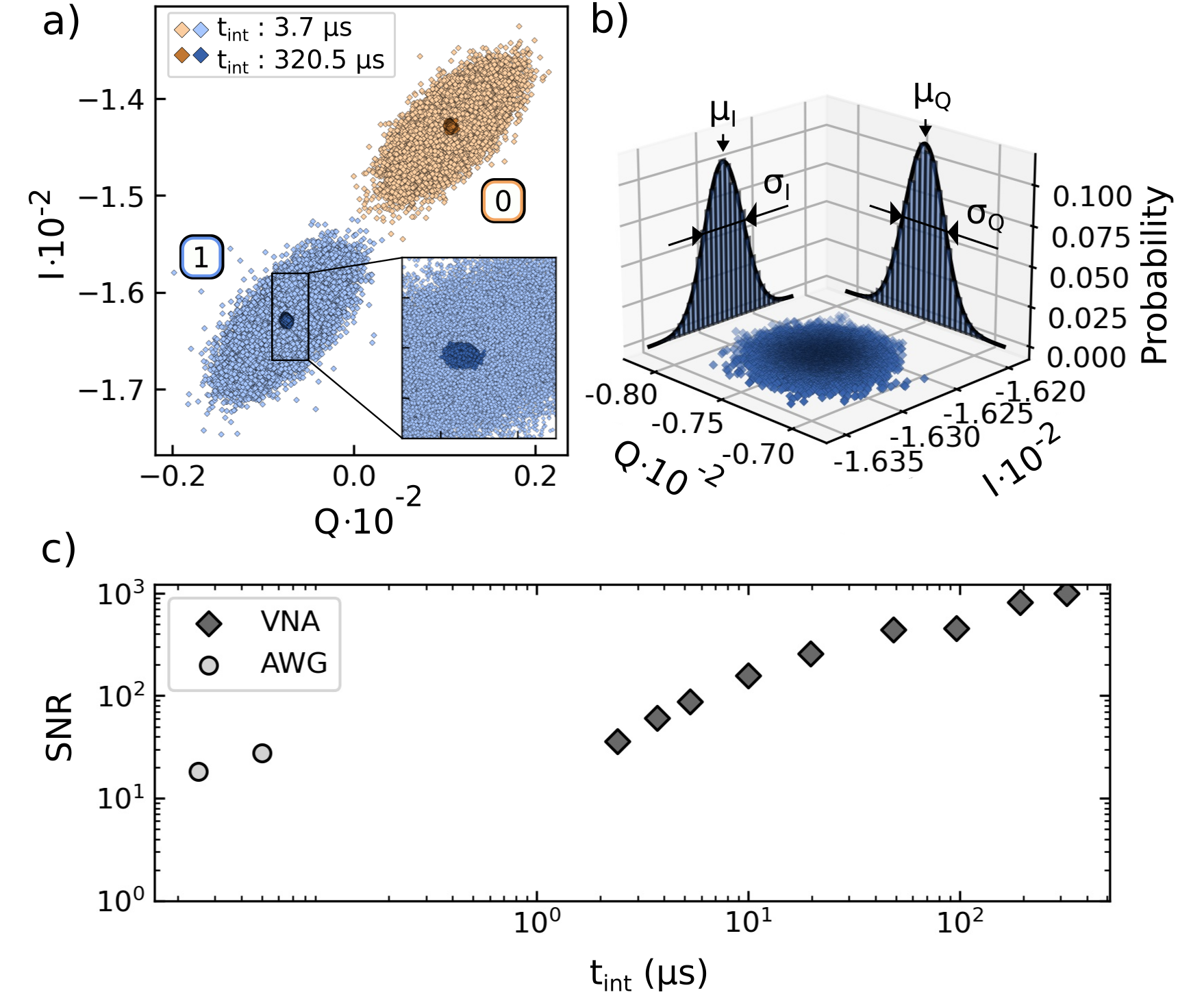}
    \caption{ a) Transmission measurements at 13 mK are represented in the $IQ$ plane. The data is collected using the VNA and $t_{int}$ = 3.7 $\mu$s (light markers) and 320.5 $\mu$s (darkmarkers). The orange markers refer to data for state 0, while the blue ones refer to state 1. b) Representative bivariate normal distribution fitting for one of the distributions presented in panel a. c) SNR as a function of $t_{\text{int}}$ for data obtained using the VNA (dark grey, diamond) and the AWG (light grey, circle).} %The blue solid line is a guide to the eye representing the SNR trend for the VNA.}
    \label{fig:3}
\end{figure}

Although the $f_r$ measurement used here represents a significant advancement over what was demonstrated for previously employed readout schemes \cite{chaves2023nanobridge,ilin2021supercurrent,ligato2021preliminary}, it remains too slow for practical applications, with every frequency sweep lasting 400 ms. Nevertheless, as shown in Fig. \ref{fig:1}b, measuring $S_{21}$ at a fixed probe frequency ($f_{probe}$) at the slope of the transmission dip close to $f_r$ is sufficient to differentiate between the two states. In this region, a small change in resonance frequency leads to a clearly detectable change in transmission. Moreover, any change in $S_{21}$ reflects a change in the in-phase ($I$) and out-of-phase ($Q$) components of the high-frequency response, with $\abs{S_{21}} = \sqrt{I^2 + Q^2}$. Measuring $I$ and $Q$ enables access to not only the magnitude of the readout high-frequency signal but also to its phase information, improving the readout resolution and making it faster and less sensitive to amplitude fluctuations \cite{Bianchetti2009dynamics}.

Each data point in Fig. \ref{fig:3}a is acquired by measuring the $IQ$ response at 13 mK at fixed $f_{probe} = 6.1381$ GHz near $f_r$ using a measurement integration time $t_{int}$ = 3.7 $\mu$s (light markers) or 320.5 $\mu$s (dark markers). At least 50000 points are collected for each $t_{int}$. During a measurement run, the memory state is switched 50 times between 0 and 1.
Thus, for each integration time, two distinct elliptical distributions appear, each corresponding to one of the two states, as represented by different colors in Fig. \ref{fig:3}a. The readout resolution can be quantified by fitting the data to bivariate normal distributions, extracting the mean values $(\mu_Q, \mu_I)$ and standard deviations $(\sigma_Q, \sigma_I)$ in both axes, as illustrated for one of the states in Fig. \ref{fig:3}b. Figure \ref{fig:3}c highlights an expected decrease of the signal-to-noise-ratio (SNR) of the VNA measurements as $t_{int}$ is shortened, with $\text{SNR} = |\mu_Q - \mu_I|/\sqrt{0.5(\sigma_Q^2 + \sigma_I^2)}$ \cite{Gambetta2007protocols,Swiadek2024enhancing}.

By extracting the readout fidelity along the quadrature axis from the overlap of the normal distributions for each state \cite{Walter2017rapid}, we obtain values better than 99.999\% for $t_{int}$ = 320.5 $\mu$s and 99.991\% for $t_{int}$ = 3.7 $\mu$s.

\begin{figure}
    \centering
    \includegraphics[width=1\linewidth]{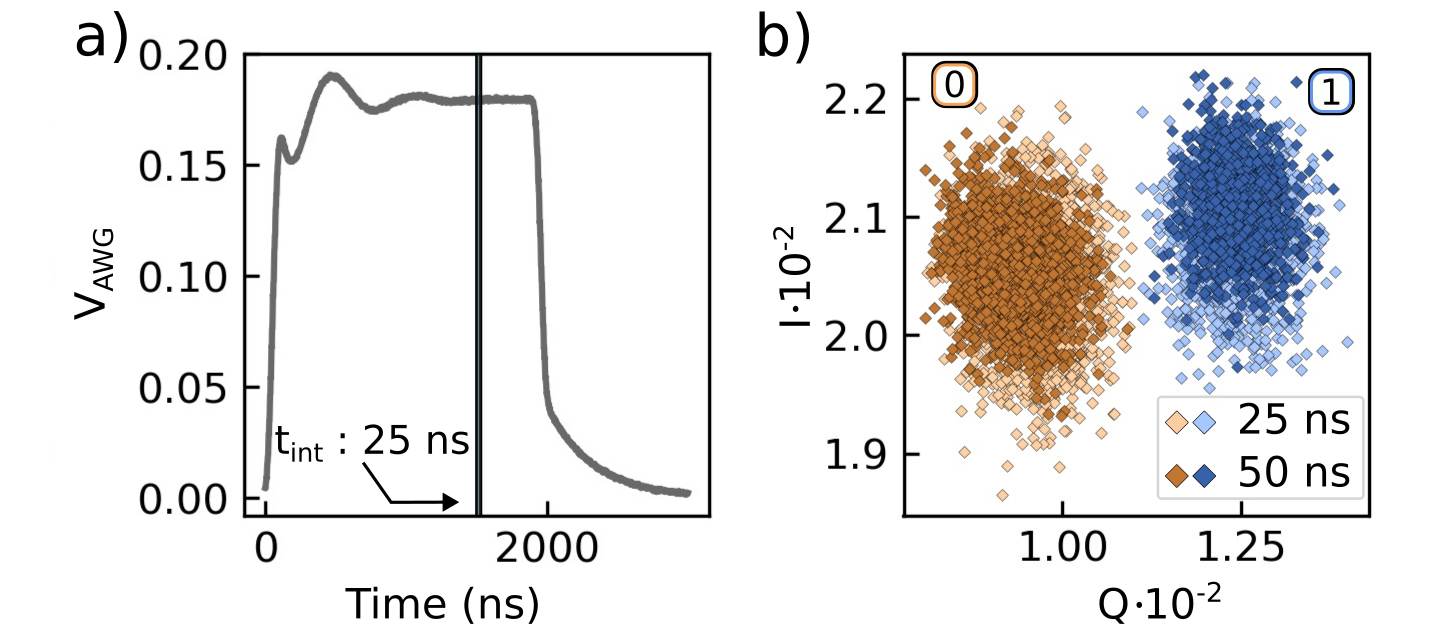}
    \caption{a) The output signal amplitude ($V_{AWG}$) of the readout pulse as a function of time. The result highlights the resonator response. The active measuring time is represented by the black vertical lines for $t_{int}$ = 25 ns. b) The acquired transmission values at 13 mK represented in the $IQ$ plane with $t_{int}$ = 25 ns (light markers) and 50 ns (dark markers) after preparing the system in states 0 and 1. State 0 is represented in orange and state 1 in blue. 
    %\lukas{change labels AWG M5300, downconverter M5201 , DAQ\_M5200, improve text size, add oscillations in the RF pulse}
    }
    \label{fig:4}
\end{figure}

Following the trend observed in Fig. \ref{fig:3}c for the VNA data, the SNR is expected to reach 1 for integration times of a few hundred nanoseconds, limiting the readout speed. This restriction could be due to the system's dynamics or to a limitation in the VNA resolution. To investigate this, we borrow concepts from cQED to perform single-shot readouts of the energy states \cite{mallet2009single,Walter2017rapid}.
%, the same procedure used for state-of-the-art qubit readout \cite{mallet2009single,Walter2017rapid}. 
This 
%approach takes advantage of the same physical principles of dispersive readout, and 
substitutes the VNA with the arbitrary waveform generator (AWG) to provide the high-frequency excitation. This equipment allows us to provide well-defined, time-resolved GHz-range pulses to probe the transmission through the feedline. The downconverter and digitizer interpret the $IQ$ response from the output signal. The measurement probe frequency can be optimized by studying the separation between the two states in the $IQ$ plane for a given $t_{int}$ \cite{krantz2019guide}. Thus, a new measurement scheme is employed for the results presented in Fig. \ref{fig:4}. First, a write pulse through the bias line selects the memory state. Then, at time zero, a 2000-ns-wide readout pulse with $f_{probe} = 6.13973$ GHz is generated by the AWG and sent through the feedline. The probe signal interacts with the system and Fig. \ref{fig:4}a shows the measured output signal amplitude, $V_{AWG} = V_{out}/V_{in}$, as a function of time at 13 mK, with the system operating under zero bias. The response demonstrates that at least 1500 ns are necessary for the resonator to reach its steady state after the start of the readout pulse. As demonstrated in Ref. \cite{jeffrey2014fast,mcclure2016rapid,Walter2017rapid}, the resonator's ring-up time can be dramatically sped up using high-power excitations, which can in principle allow for settling times faster than 100 ns.

Taking the resonator dynamics into account, we actively probe the sample for a specific $t_{int}$ after a 1500 ns delay. This process is repeated for at least 1500 high-frequency readout pulses in both state 0 and state 1. Figure \ref{fig:4}b shows the resulting data in the $IQ$ plane for $t_{int}$ = 25 ns and 50 ns, with their corresponding SNR values included in light gray circles in Fig. \ref{fig:3}c. This comparison shows that the AWG-based measurement drastically outperforms the expected SNR of the VNA at low integration times. The readout fidelity achieved with the AWG is 99.983\% and 99.698\% for 50 ns and 25 ns, respectively. 
As such, the memory dynamics and the spacing between different states in the $IQ$ plane allow us to push readout times to a few hundreds or possibly tens of nanoseconds. When paired with the possibility of fast write operations, this fact demonstrates that the proposed phase slip memory can be operated at very high speeds. It is worth noticing that the presented measurement scheme allows single-shot resolution using a 25-ns sampling although the shift in resonance frequency between the two states at zero bias is only about 10 kHz.

In summary, we have fabricated a nonvolatile classical memory element that stores information in the winding of a high-kinetic inductance Al loop. The fabrication process is simple and the loop's geometry offers a control knob to the device's energy landscape, allowing to tune its switching dynamics. The memory states are reliably manipulated via an on-chip current bias line, which generates a local magnetic field that directly influences the loop, eliminating the need for external magnetic field sources—a key advantage for integration with other on-chip components.
By coupling the Al loop to a NbTiN superconducting coplanar waveguide resonator, we perform energy-efficient readout operations with a 99.698\% fidelity for an active sampling time of 25 ns during single-shot measurements.
The memory element realizes the full potential of phase-slip-based classical memories, combining their inherent stability with state-of-the-art readout capabilities and straightforward control. Additionally, incorporating cQED concepts into the memory design and operation ensures intrinsic compatibility with modern quantum information systems. This approach could be extended to the design and integration of phase-slip qubits, a goal that appears achievable by simply modifying the loop's geometry and tuning its characteristic energy barrier \cite{mooij2005phase, peltonen2013coherent, Randeria2024dephasing,Honigl-Decrinis2023}.

% \section*{Data availability}
% The data that support the findings of this study are available from the corresponding 
% authors upon reasonable request.

\section*{Acknowledgments}
The authors would like to thank Ekaterina Gorokh for her help during the Al growth procedure. L.N acknowledges support from FWO research fellowship nr. 11K6525N. S.R. acknowledges support from FWO research fellowship nr 11A3V25N. This research is supported and funded by a Flemish interuniversity BOF project (IBOF-23-065). The authors acknowledge the financial support of the KU Leuven C18/24/110 program and the FWO-FNRS Weave program. NbTiN resonators were fabricated in the Myfab cleanroom, Chalmers.

\section*{Author contributions}
L.N., S.R., and I.P.C.C. fabricated the devices. L.N. and R.D. executed the measurements. D.A.D.C., B.R., M.J.V.B. and J.V.V. conceptualized the experiments. K.T., B.R., M.J.V.B., and J.V.V. supervised the research. L.N., D.A.D.C., and J.V.V. wrote the original manuscript. All authors discussed the results and reviewed the final manuscript. 

% \section*{Competing interests}
% The authors declare no competing interests.

\bibliography{Ref.bib}
\end{document}

\title{Supplementary Information to ``Nonvolatile Cryogenic Phase Slip Memory with Single-Shot Readout''}

\author{Lukas Nulens}
    \email{lukas.nulens@kuleuven.be}
    \affiliation{Quantum Solid-State Physics, Department of Physics and Astronomy, KU Leuven, Celestijnenlaan 200D, Leuven, B-3001, Belgium}

\author{Davi A. D. Chaves}
    \affiliation{Quantum Solid-State Physics, Department of Physics and Astronomy, KU Leuven, Celestijnenlaan 200D, Leuven, B-3001, Belgium}

\author{Stijn Reniers}
    \affiliation{Quantum Solid-State Physics, Department of Physics and Astronomy, KU Leuven, Celestijnenlaan 200D, Leuven, B-3001, Belgium}

% \author{Ekaterina Gorokh}
%     \affiliation{Quantum Solid-State Physics, Department of Physics and Astronomy, KU Leuven, Celestijnenlaan 200D, Leuven, B-3001, Belgium}

\author{Ruben Dillemans}
    \affiliation{Quantum Solid-State Physics, Department of Physics and Astronomy, KU Leuven, Celestijnenlaan 200D, Leuven, B-3001, Belgium}

\author{Ivo P. C. Cools}
    \affiliation{Quantum Device Physics Laboratory,
Department of Microtechnology and Nanoscience,
Chalmers University of Technology, Kemivägen 9, Goteborg, SE-412 58, Sweden}

\author{Kristiaan Temst}
    \affiliation{Quantum Solid-State Physics, Department of Physics and Astronomy, KU Leuven, Celestijnenlaan 200D, Leuven, B-3001, Belgium}
\author{Bart Raes}
\affiliation{Imec, Kapeldreef 75, 3001 Leuven, Belgium}
\author{Margriet J. Van Bael}
    \affiliation{Quantum Solid-State Physics, Department of Physics and Astronomy, KU Leuven, Celestijnenlaan 200D, Leuven, B-3001, Belgium}

\author{Joris Van de Vondel}
    \email{joris.vandevondel@kuleuven.be}
    \affiliation{Quantum Solid-State Physics, Department of Physics and Astronomy, KU Leuven, Celestijnenlaan 200D, Leuven, B-3001, Belgium}

%\date{\today} % Leave empty to omit a date

\maketitle

\begin{figure}[H]
    \centering
    \includegraphics[width=1\linewidth]{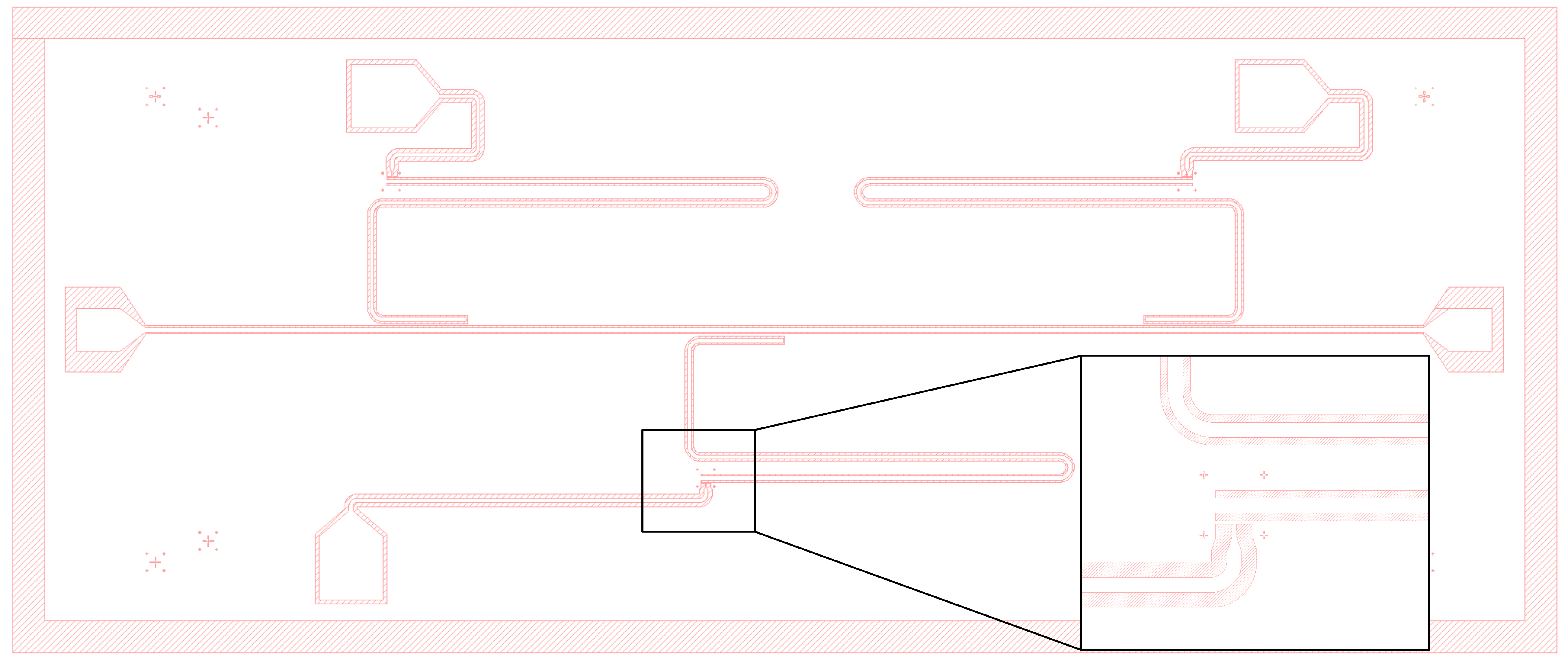}
    \caption{An overview of the sample design shows three $\lambda$/4 resonators capacitively coupled to a central feedline, with white representing NbTiN and red indicating the etched areas. A control is implemented at each resonator, as shown in the inset. An Al loop is placed in the trench of each resonator. The loops were designed with varying dimensions to tune the energy barrier to the desired range. For two of the loops, the energy barrier was too high to induce a phase slip at the maximum applicable current. The remaining loop, which met the required conditions, is discussed in the main manuscript.}
    \label{fig:2}
\end{figure}

\begin{figure}[H]
    \centering
    \includegraphics[width=1\linewidth]{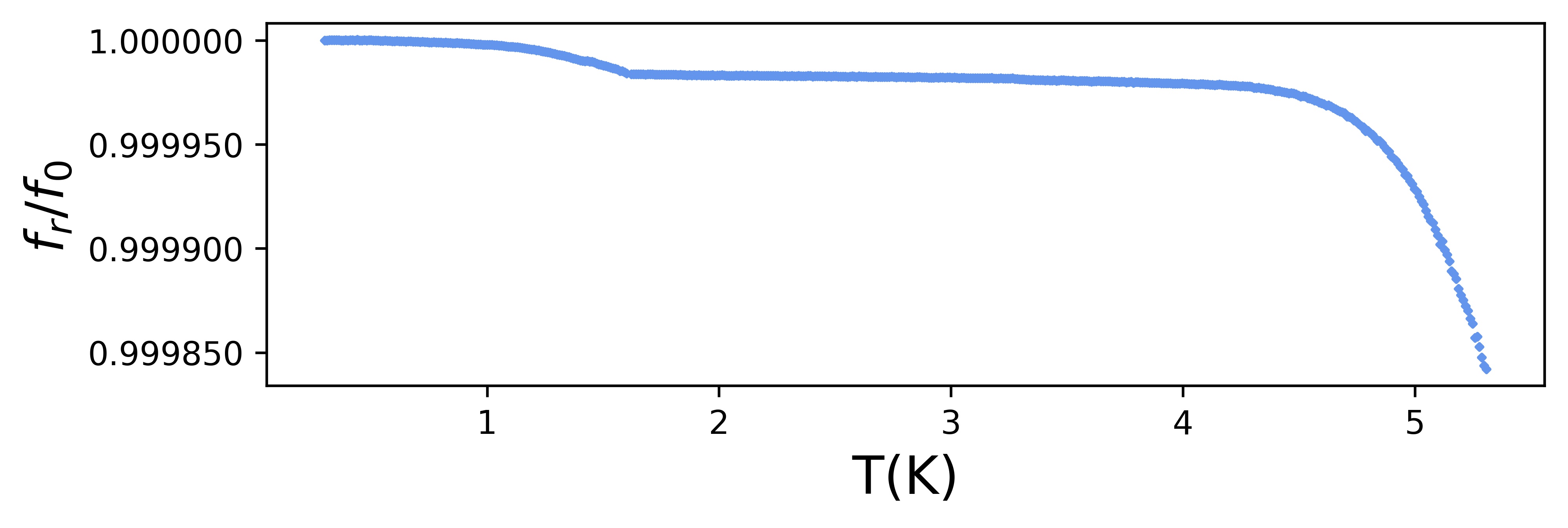}
    \caption{The resonator's resonance frequency, normalized to its 300 mK value $f_0=f_r(T=300\,\text{mK})$, as a function of temperature. Around 1.6 K, a bump in $f_r(T)$ signals the Al loop’s transition to the superconducting state. This transition introduces a kinetic inductance contribution, altering the resonator's frequency through mutual inductive coupling, as discussed in Ref. \cite{nulens2024noninvasive}. At higher temperatures ($>4$ K), changes in NbTiN's kinetic inductance also affect the resonance frequency.
}
    \label{fig:2}
\end{figure}

\begin{figure}[H]
    \centering
    \includegraphics[width=1\linewidth]{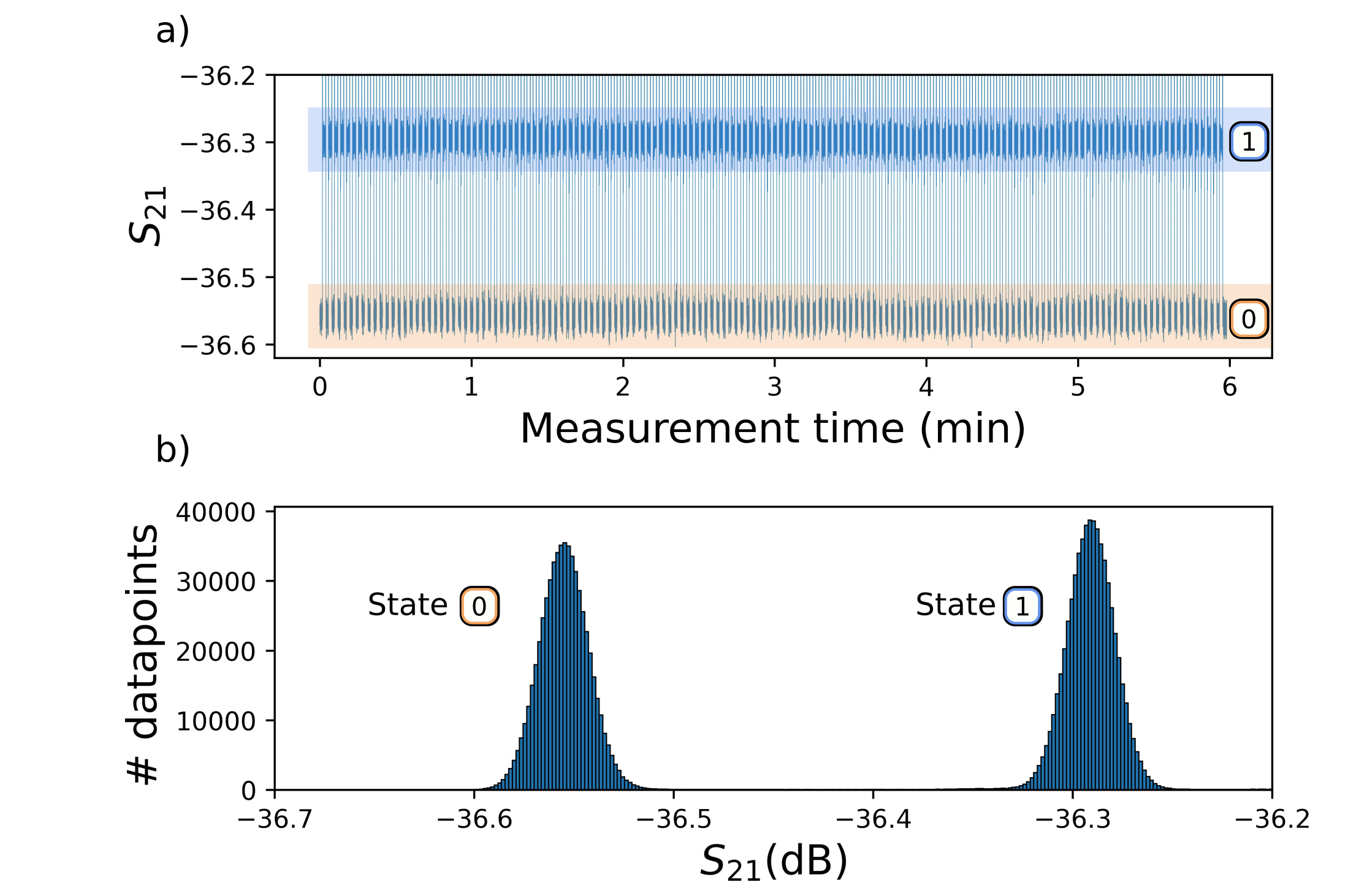}
    \caption{Reliability of the write operation. a) The transmission through the feedline at a fixed frequency $f_{probe} = $ 6.1380842 GHz as a function of time. During the time showed in the panel, the state is switched a total of 300 times between states 0 and 1. This experiment was continuously repeated during 3 hours, with 6000 write operations switching the state. After all pulses, the state is correctly switched. Panel b) shows a distribution of the transmission measurements after each pulse for all 6600 write operations, confirming no spurious states are observed.}
    \label{fig:2}
\end{figure}

\begin{figure}[H]
    \centering
    \includegraphics[width=1\linewidth]{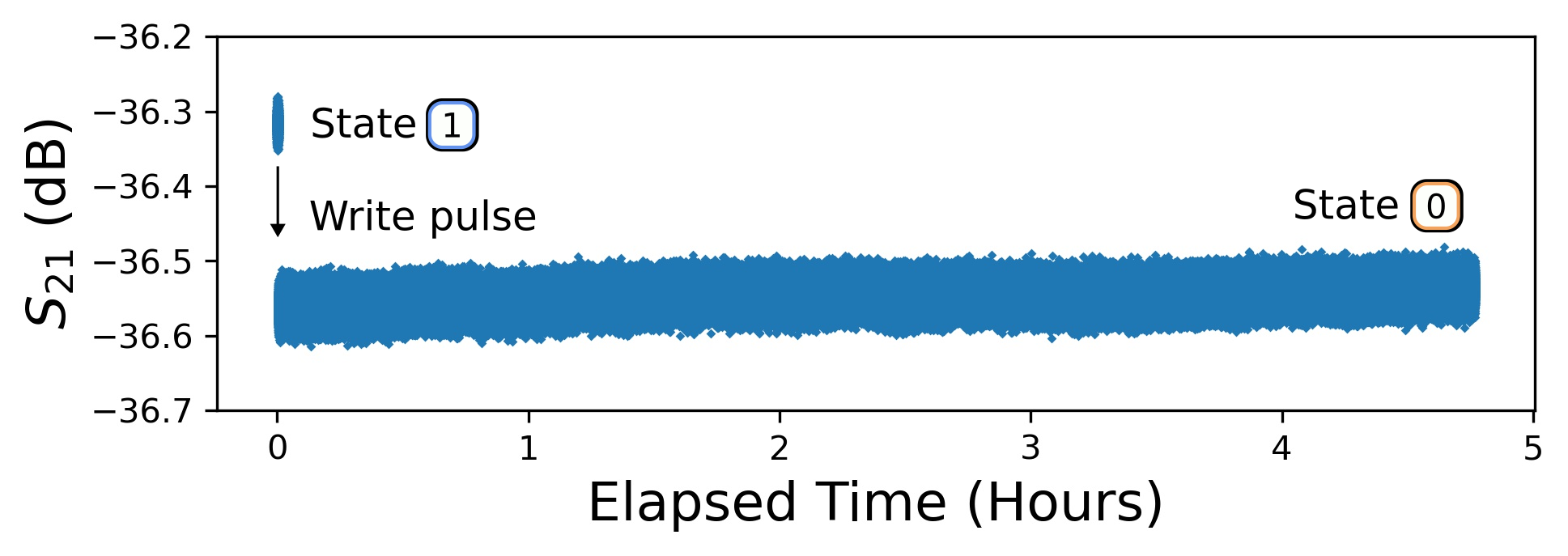}
    \caption{ State stability check under zero external bias. The figure shows the transmission through the feedline at a fixed frequency $f_{probe} = $  6.1380842 GHz as a function of time. At the start of the measurement, the loop is found in state 1. Then a pulse causes the loop to transition to state 0. From that moment, the transmission was continuously monitored for approximately 5 hours, during which time the loop remained stable in the prepared energy state.
}
    \label{fig:2}
\end{figure}

\bibliography{Ref.bib}